# Strain-induced magnetization control in an oxide multiferroic heterostructure

*Federico Motti*[1, 2], Giovanni Vinai[1], Aleksandr Petrov[1], Bruce A. Davidson[1], Benoit Gobaut[3], Alessio Filippetti[4,5], Giorgio Rossi[1, 2], Giancarlo Panaccione[1], Piero Torelli[1]*

[1] Laboratorio TASC, IOM-CNR, S.S. 14 km 163.5, Basovizza, I-34149 Trieste, Italy

[2]Dipartimento di Fisica, Università degli Studi di Milano, Via Celoria 16 I-20133 Milano, Italy

[3]Elettra Sincrotrone Trieste S.c.P.A., SS 14 km 163.5, Basovizza, I-34149 Trieste, Italy

[4] IOM-CNR, S.P. Monserrato-Sestu km 0.7, Monserrato (CA), I-09042, Italy

[5]Dipartimento di Fisica, Università di Cagliari, S. P. Monserrato-Sestu Km.0.7, Monserrato (CA), I-09042, Italy

**Abstract**

Controlling magnetism by using electric fields is a goal of research towards novel spintronic devices and future nano-electronics. For this reason, multiferroic heterostructures attract much interest. Here we provide experimental evidence, and supporting DFT analysis, of a transition in $La_{0.65}Sr_{0.35}MnO_3$ (LSMO) thin film to a stable ferromagnetic phase, that is induced by the structural and strain properties of the ferroelectric $BaTiO_3$ (BTO) substrate, which can be modified by applying external electric fields. X-ray Magnetic Circular Dichroism (XMCD) measurements on Mn L edges with a synchrotron radiation show, in fact two magnetic transitions as a function of temperature that correspond to structural changes of the BTO substrate. We also show that ferromagnetism, absent in the pristine condition at room temperature, can be established by electrically switching the BTO ferroelectric domains in the out-of-plane direction. The present results confirm that electrically induced strain can be exploited to control magnetism in multiferroic oxide heterostructures.

## I. INTRODUCTION

After revolutionizing the data storage technology, the control on the electron spin is now close to be implemented in nanotechnology for applications in computation, communication and energy harvesting[1,2]. To realize such innovative spintronic devices, one of the key challenges is to find reliable, fast and energy efficient ways to manipulate the magnetic state in a material or heterostructure. Controlling (ferro)magnetism via application of an electric field appears very attractive as no large power-dissipating currents[3,4] are needed in principle. Electric field control of magnetism has been obtained in multiferroics[5] but they usually display a weak ferromagnetic response[6,7]. To overcome this limitation, the use of artificial heterostructures combining ferromagnetic films with ferro- or piezo-electric substrates has been explored[8–12].

Transition metal oxides with perovskite structure are promising in this context, as they display strong correlation between spin, charge, orbital and lattice degrees of freedom, thus potentially providing multiple ways to influence magnetism[13–15]. It has been previously shown that the total magnetic moment[16,17], the coercive field[18], the magnetic anisotropy[19,20] and the Curie temperature[17,21], can be modified by applying electric fields to oxide heterostructures. A magnetic transition within the thickness of few unit cells driven by charge accumulation at the manganite/ferroelectric interface was also demonstrated[22]. However, the possibility use strain to reversibly drive a magnetic transition on a longer scale is still worth to be explored, both for fundamental scientific interest and potential practical applications.

We show here the results of element-specific magnetometry on $BaTiO_3/La_{1-x}Sr_xMnO_3$ (BTO/LSMO) epitaxial heterostructure grown by Molecular Beam Epitaxy, as a function of the modified strain of the substrate, which reversibly triggers phase transitions in the LSMO overlayer. The applied strain is tuned employing the intrinsic structural transition of the substrate for changing temperature, as well as switching its ferroelectric domains with an electric bias. The main result is the development of ferromagnetism at 300 K in LSMO driven by BTO poling.

The phase diagram of ferroelectric BTO displays four crystal structures, that are stable at different temperatures[23,24]. The rhombohedral (R, below 180 K), orthorhombic (O, between 180 and 280 K) and tetragonal (T, up to 410 K) phases are all ferroelectric, with the polarization vector pointing along [111], [011] and [001] pseudo-cubic directions, respectively. A structural phase transition into a cubic, non-ferroelectric lattice, takes place at 410 K. LSMO presents a complex phase diagram, displaying ferromagnetic as well as various kind of anti-ferromagnetic order depending of temperature and La/Sr ratio[25]. The magnetic state of LSMO is reflected in its transport properties.[26] For the Sr-doping concentration $x$=1/3, bulk LSMO is ferromagnetic and metallic with Curie temperature above room temperature (around 370 K). However, the physical properties of LSMO can be tuned also by means of epitaxial strain,[27,28] and are therefore affected by the BTO structural phase and polarization orientation.[19]

We have grown ultrathin (30 u.c. ≈ 12 nm) films of LSMO by UHV Molecular Beam Epitaxy on top of a BTO substrate obtaining fully epitaxial heterostrcture, as demonstrated by RHEED images acquired during the deposition. We have probed the magnetic properties of the LSMO by measuring X-ray Magnetic Circular Dichroism (XMCD) on the Mn $L_{2,3}$ edge at the APE-HE beamline of the Elettra synchrotron radiation facility in Trieste.[29,30] LSMO is observed to undergo magnetic transitions when changing the temperature and, at 300 K, when applying electric bias. X-ray diffraction (XRD) measurements in Bragg-Brentano geometry show that these effects on the overlayer are correlated to the structural changes of the BTO substrate, i.e. are connected with modifications of the interface constraints. Ab-initio Density Functional Theory (DFT) simulations, as implemented with the QUANTUM ESPRESSO code, have been performed giving independent support of the reproducible observation of strain-mediated magnetic transitions in the LSMO layer.

## II. EXPERIMENTAL RESULTS

We concentrate the analysis on LSMO thin films (doping level $x$=0.35, thickness 30 u.c.) grown epitaxially on a BTO crystalline substrate (thickness 1 mm).

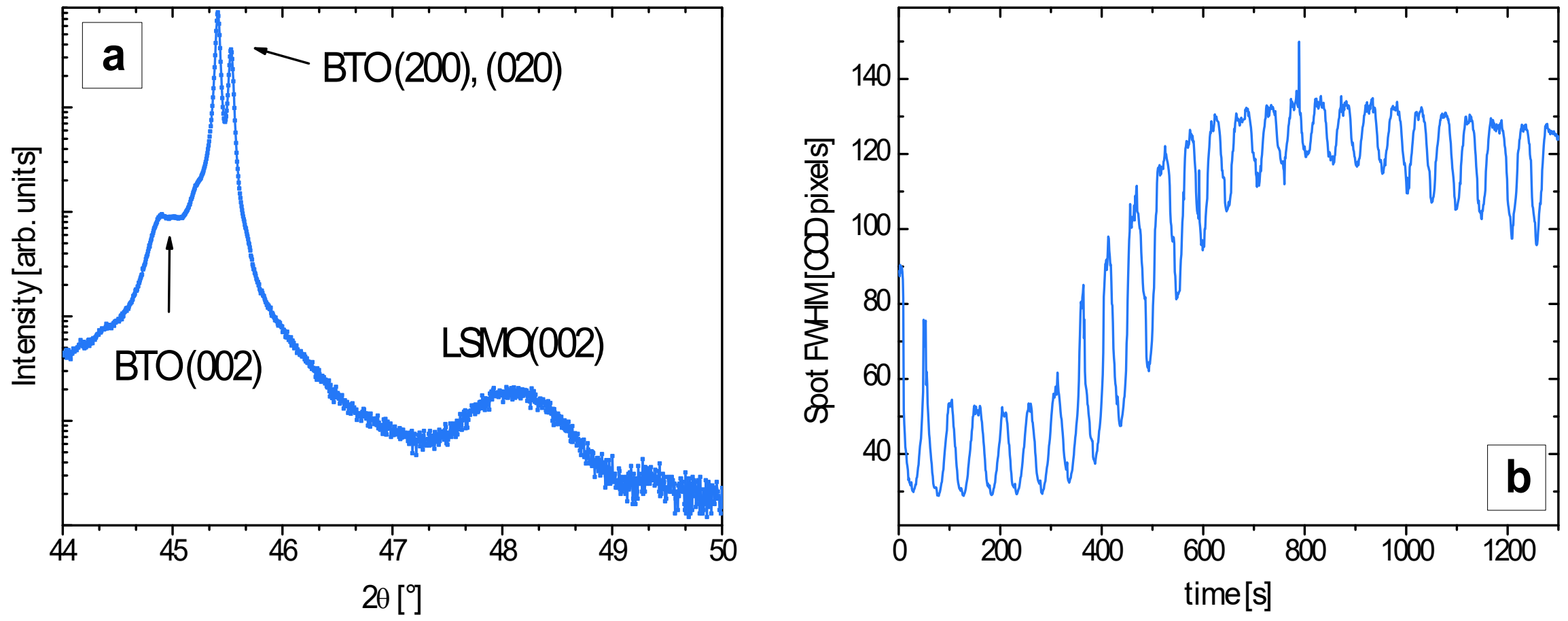

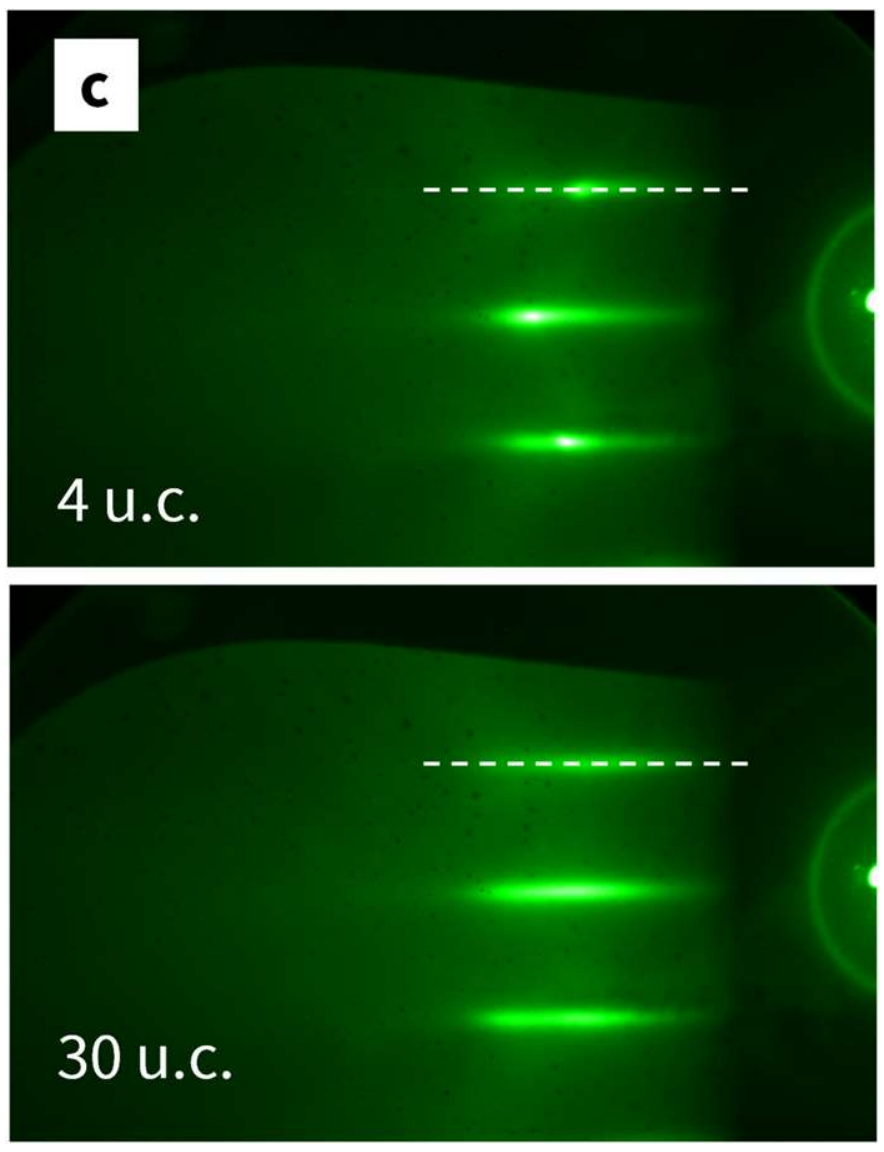

**Figure 1 – (a)** θ-2θ scan of LSMO/BTO sample, acquired at room temperature in a high-intensity configuration. It is possible to distinguish the different domains of BTO crystal in the tetragnal phase, both out-of-plane and in-plane. Note that BTO peaks are split into two components because of the presence of both $K_{\alpha1}$ and $K_{\alpha2}$ lines in the Cu X-ray source. **(b)** Evolution of the full width half maximum of RHEED diffraction spots during the growth of LSMO, starting from the 5th unit cell. **(c)** RHEED images acquired in situ during LSMO growth, after completing the $4^{th}$ and the $30^{th}$ unit cell. The plot in Figure 1b was obtained from the profile along the dashed lines.

**Figure 1a** shows the XRD θ-2θ diffraction scan in high intensity mode at room temperature for as-deposited LSMO/BTO. Non-polarized BTO shows the expected presence of domains elongated both in-plane, i.e. (100) and/or (010), and out-of-plane, (001), in T phase. The corresponding calculated lattice parameters of BTO are 3.991 Å and 4.035 Å respectively, in perfect agreement with the data reported in literature.[23,24] By comparing the relative intensities of the in-plane and out-of-plane peaks, we infer that the majority of domains are oriented in-plane.

Regarding the LSMO thin layer, its (002) peak indicates a pseudo-cubic out-of-plane lattice parameter of 3.78 Å, much smaller compared to the bulk value of 3.87 Å.[31] This is due to the substrate-induced in-plane tensile strain, which causes a decrease of the out-of-plane lattice parameter. The full-width-half-maximum (FWHM) of the RHEED (01) diffraction spot, shown in **Figure 1b**, was recorded to monitor the dynamics of the crystalline structure of the film. A broadening of the diffraction spots is observed after 10 u.c., symptom of the increasing disorder originating from the formation of defects and/or surface roughening. A further effect to be considered is the domain structure and mosaicity of the substrate. The formation of defects for increasing thickness can be expected given the large mismatch (2.6% - 3.3%, depending on the structural phase) between BTO and LSMO, and may accompany the tendency to change the lattice parameters towards bulk values (relaxation). However, given the value of the out-of-plane lattice parameter measured, the film appears to be far from the fully-relaxed bulk structure, and still clamped to the substrate. Using the poisson ratio $\nu=0.36$ reported in literature[32], an expanded in-plane lattice parameter of 3.90 Å is calculated. A reciprocal space map around the (103) reflection is presented and discussed in the Supporting Information[33]. These data testify a partial relaxation of the LSMO film. We notice that the (103) reflection of the film is very low, possibly because of the poor quality of the BTO substrate.

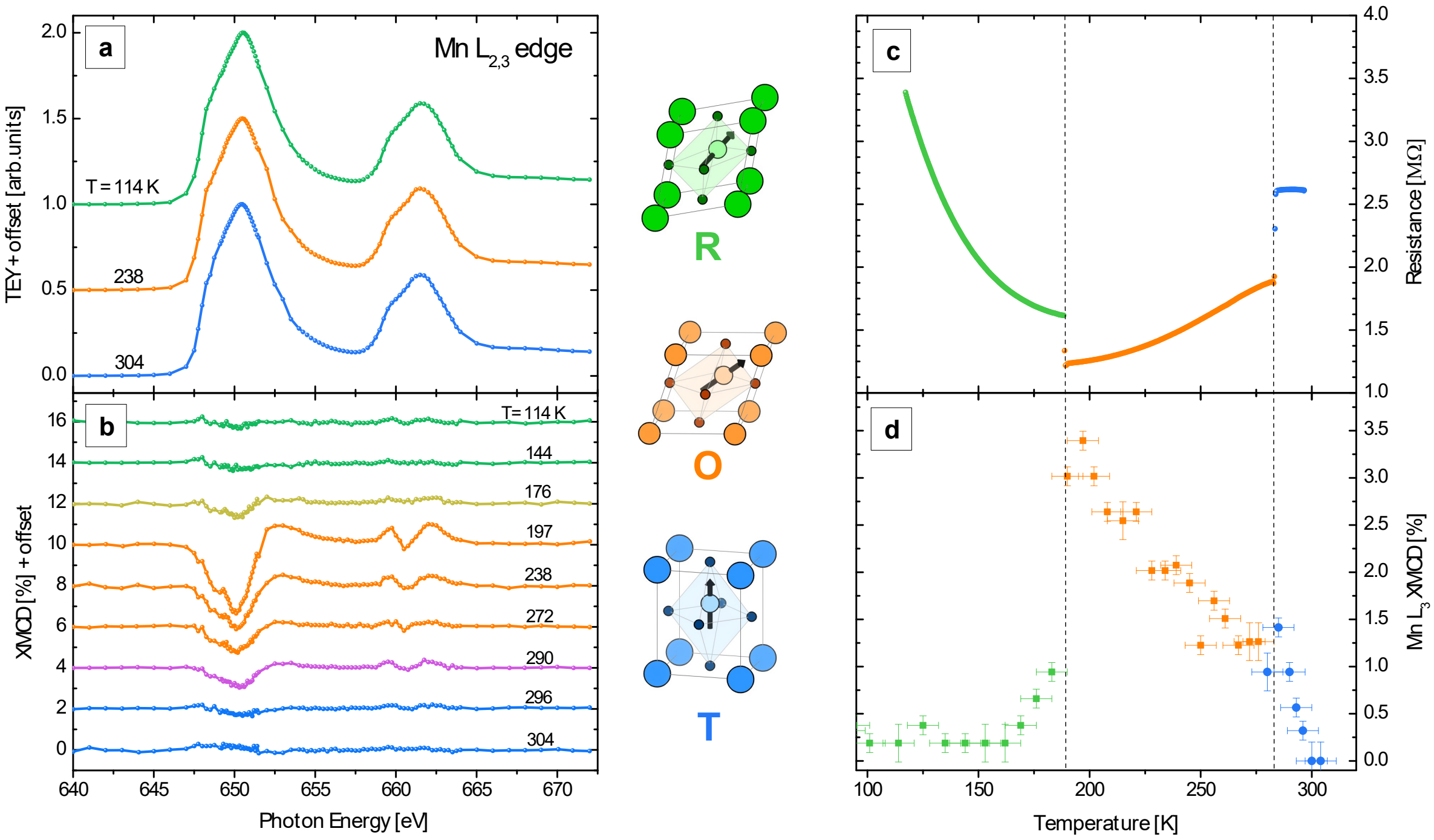

**Figure 2 –** Mn $L_{2,3}$ XAS (a) and XMCD (b) spectra acquired for various temperatures corresponding to different BTO structural phases, for the pristine case. The XAS curves shown are the sum of the two absorption spectra measured after magnetic field saturation with opposite field directions. Temperature dependence of LSMO/BTO resistance (c) and XMCD signal on the Mn $L_3$ edge (d) with BTO in the pristine state. Dashed lines correspond to BTO structural transitions.

**Figure 2a, 2b** show the absorption spectra and corresponding XMCD curves of the unpolarized LSMO/BTO sample. The XMCD values expressed in % have been corrected taking into account the angle of 45° between the incident beam light and the direction of the in-plane applied magnetic field, as well as the 75% circular polarization degree of our undulator light.

The XAS spectrum presents two broad multiplets, due to the large Mn 3*d* bandwidth, as expected and previously reported for optimally doped LSMO.[34–36] When passing across the BTO structural transitions, no changes were observed in the Mn $L_{2,3}$ lineshape, as shown in Figure 2a. However, a clear change was observed in the corresponding dichroism, as shown in Figure 2b: for the BTO rhombohedral (T < 180 K, in green) and tetragonal (T > 280 K, blue) phases no dichroism was detected in the LSMO overalyer, but in the orthorhombic phase (orange) a XMCD signal of 3% is clearly detected. The measured multiplet structure corresponds to what reported in literature for optimally doped LSMO thin films.[17,37] These results show that even if the structural phase of the BTO substrate do not modify the chemical environment of Mn in LSMO, it does affect its magnetic ordering.

**Figure 2c** shows the electric transport measurements of LSMO/BTO in the temperature range between 120 and 300 K obtained with the four-probes method in Van der Pauw configuration. We observe jumps of resistance values in correspondence of all the BTO structural transitions. Such sharp transitions were also reported for thicker LSMO layers on BTO.[19] The transport properties are well correlated with the magnetic changes observed with XMCD. In the O phase the resistance increases with temperature, which is typical of a metallic behavior, whereas in the R phase it decreases, as expected for a semiconductor/insulator. It is known that in LSMO there is a strong connection between electric transport and magnetic ordering, due to the double-exchange mechanism, so that ferromagnetism is related to a metallic phase whereas the insulator behavior is a sign of lack of ferromagnetic order.[38] This is confirmed also in our case, with a perfect correlation between transport and XMCD measurements (**Figure 2d**). The LSMO magnetic transitions measured in correspondence of the structural transitions of BTO proved to be perfectly reproducible and independent of the thermal history of the sample. Consistent data were measured during the cooling of the sample.

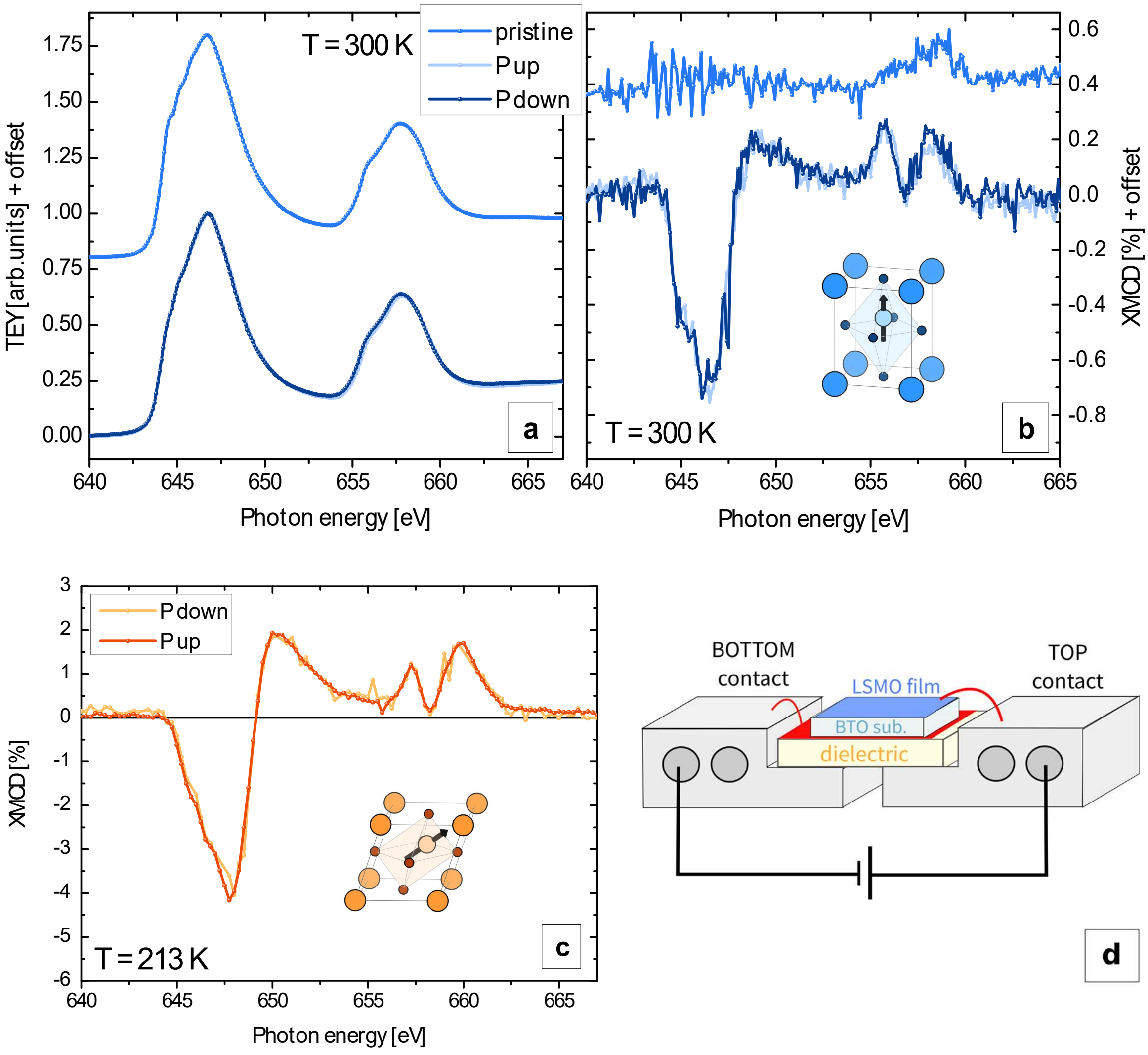

**Figure 3** –Comparison of the Mn $L_{2,3}$ XAS (a) and XMCD (b) spectra for BTO in the pristine state and polarized with positive or negative bias in T phase. (c) Comparison of the XMCD spectra for BTO polarized with positive/negative bias, in the O phase. (d) Schematic of the sample holder used for in-situ polarization of BTO. Contacts with the sample holder (in red) were made with silver paint. The dielectric spacer was inserted to avoid shorts between the two parts of the sample holder.

Upon out-of-plane polarization of BTO at room temperature, a similar evolution of the XMCD signal with temperature was observed: no dichroism was detected in the lowest temperature range (BTO in R phase) but a clear signal of magnetic dichroism was detected for BTO in the O phase. This XMCD signal was measured also without external magnetic field by reversing the light circular polarization handedness, as well as when the sample reached this state being warmed up from the non-ferromagnetic R phase. This shows that the LSMO film acquires a spontaneous remanent magnetization after the BTO R/O phase transition.

A smaller but clearly detected XMCD signal, in the range 0.5-1%, was also measured at room temperature, which was absent in the pristine non-polarized system (**Figure 3b**). This variation in the LSMO magnetization is again not reflected in changes in the XAS lineshape, as shown in **Figure 3a**. No differences could be detected in the spectra when reversing the direction of the polarization for all the BTO structural phases. This was verified both at room temperature (as shown in Figure 3b) and with BTO in the O phase, for which the highest dichroic signal is observed (**Figure 3c**). The effective change of the polarization state was monitored acquiring a current vs. voltage curve (see supplementary information). The unchanging XAS/XMCD spectrum is compatible with entirely strain-driven magnetic phenomena, and excludes charge accumulation/depletion effects at the BTO-LSMO interface as a possible origin.

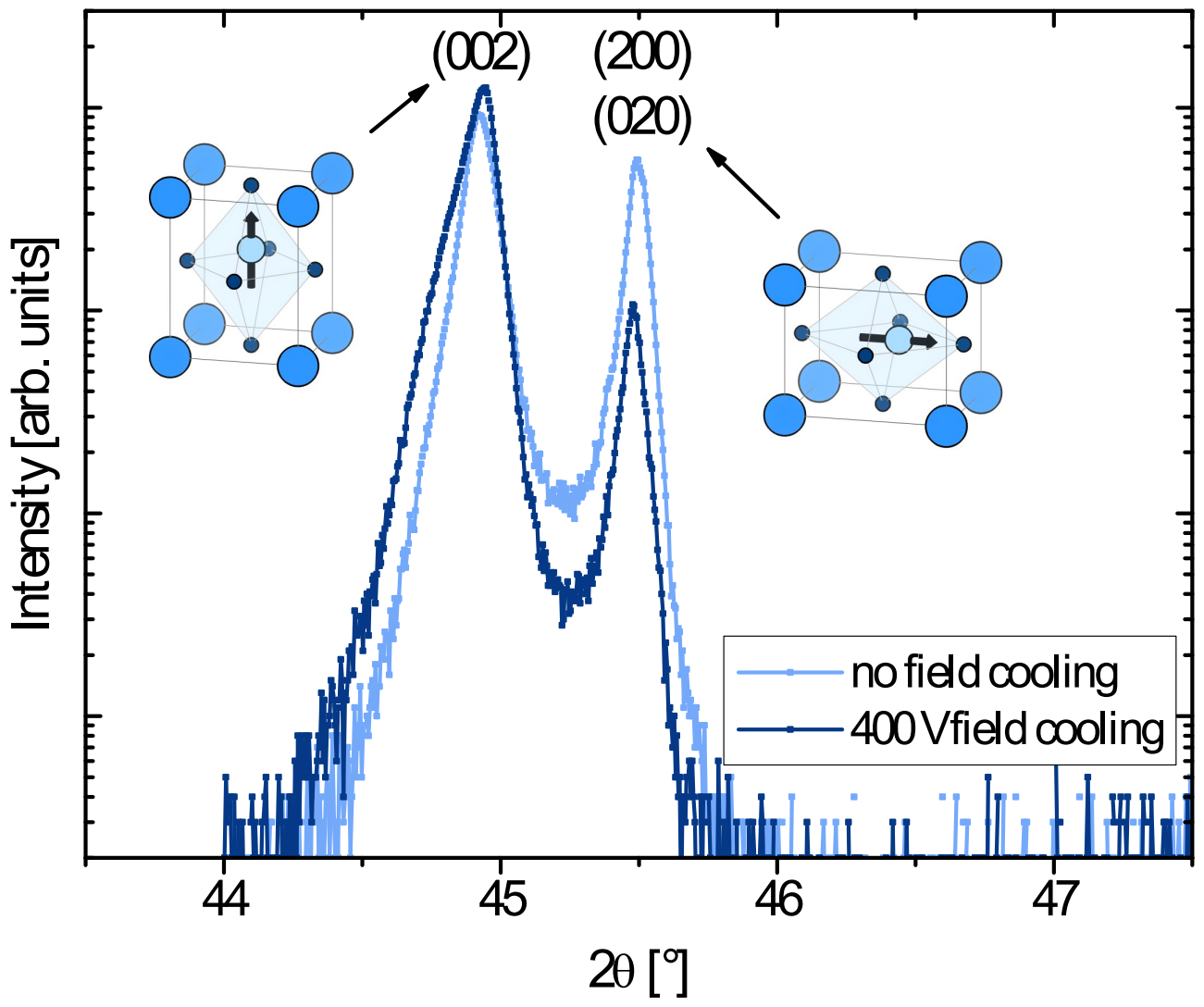

**Figure 4 –** θ-2θ scans in high-resolution configuration of LSMO/BTO sample without field cooling from the cubic phase (light blue) and after field cooling under applied 400 V (dark blue). On the side, schematics of BTO unit cells for (001) and (010).

In order to observe the structural variations of BTO after setting the out-of-plane electric polarization, HR-XRD θ-2θ scans of LSMO/BTO were performed. Firstly, the sample was set in the high temperature cubic phase, then cooled down to room temperature (tetragonal phase); the measurements were performed both without applied bias voltage (light curve in **Figure 4**) or with an out-of-plane applied electric field of 400 V (dark curve). Unpolarized BTO presents a combination of in-plane (100) and (010) and out-of-plane (001) domains, as sketched in the insets of Figure 4. When an electric field is applied along the *c*-axis (perpendicular to the surface), BTO aligns its dielectric polarization which implies shrinking the in-plane lattice parameter and expanding the out-of-plane one. The ratio between the two domains changes consequently, and most domains are set in the (001) direction: the θ-2θ scans show a dramatic change in the out-of-plane/in-plane peak heights, which is compatible with the out-of-plane rotation of the ferroelectric domains. The same effect is expected to occur when applying a voltage at a fixed temperature, consistently with previous observations by Eerenstein *et al.*[16]. This was done during our XMCD measurements.

## III. AB-INITIO CALCULATIONS

DFT calculations of strained LSMO were performed in order to gain a better understanding of the complex observed phenomenology. A $\sqrt{2}\times\sqrt{2}\times2$ cell with tetragonal/orthorhombic *Pnma* symmetry was assumed, with generic $a^-b^-c^+$ octahedral tilting pattern (this symmetry also characterizes the AFM $LaMnO_3$ endpoint structure). In the simulations, the interface plane lattice parameters *a* and *b* (either square or rectangular) were fixed, while the system was fully relaxed along the interface-perpendicular direction (*c* axis). A tight convergence threshold of 0.1 mRy/Bohr was imposed to the forces. As for magnetic ordering, we considered FM ordering and three different AFM ordering, i.e. A-type, C-type, and G-type; in this way the nearest-neighbor interactions along all the three directions was taken into account. The Sr doping level is 25% in all the calculations presented hereafter.

Two sets of simulations were performed: in the first set a squared substrate, i.e. with $a = b$, was imposed; this mimics LSMO grown on BTO at room-temperature when polarized out-of-plane. In the second set we allowed $a \neq b$ to explore a possible tetragonal-to-orthorhombic symmetry lowering for LSMO. This could mimic the distortion imposed by the BTO substrate in correspondence with the transition from the T to the O phase. However, structural disorder and/or configurational entropy effects are not included in the supercell approach.

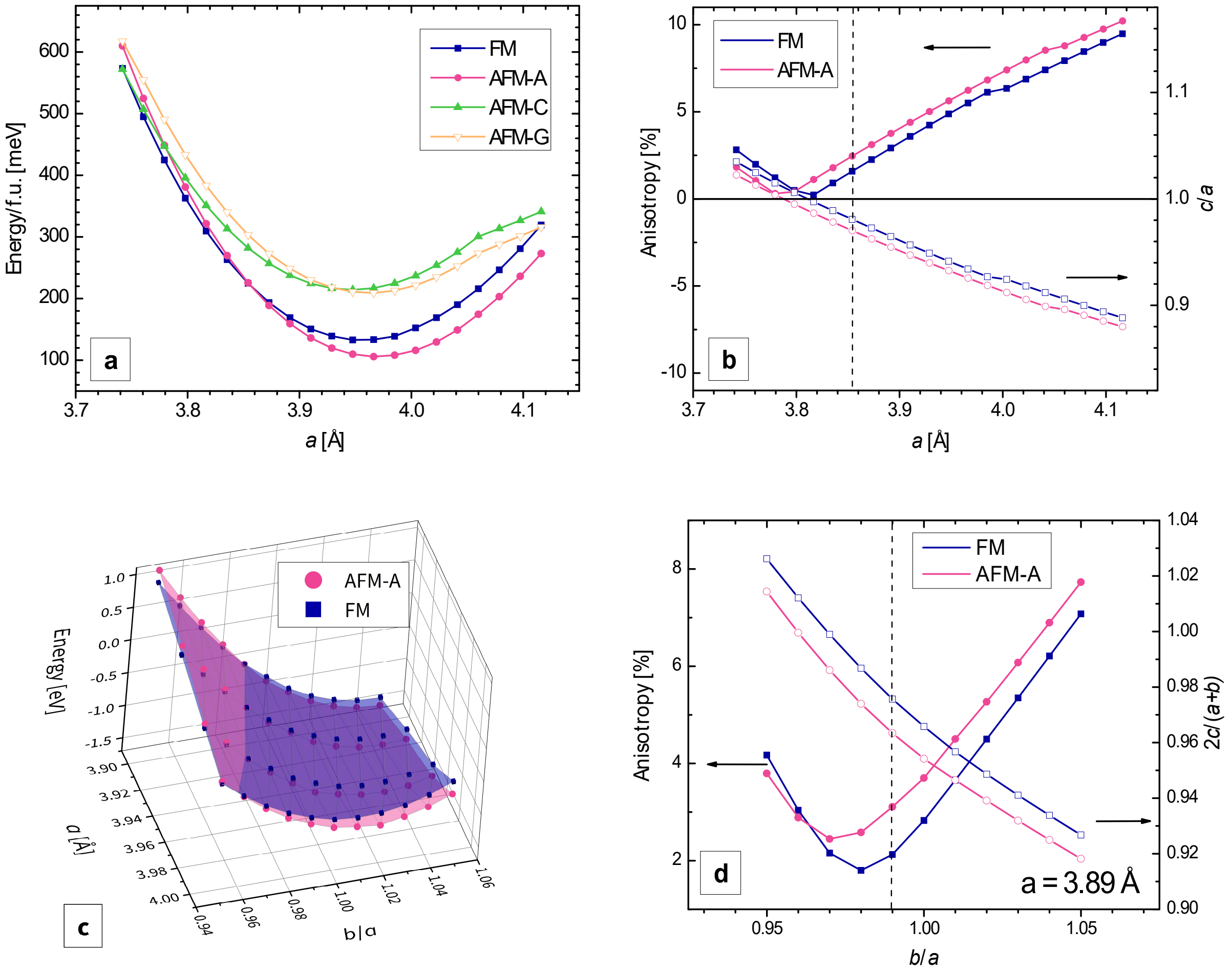

**Figure 5** - Calculations of tetragonal LSMO film under planar strain: (a) energy per formula unit; (b) anisotropy factor (left axis) and *c/a* ratio (right axis). Calculations of orthorhombic *Pnma* LSMO: (c) energy per cell as a function of both a and b/a; (d) example of anisotropy factor (left axis) and *c/a* ratio (right axis) for fixed *a*=3.89 Å. In (b) and (d) the dashed lines correspond to the *a* and *b/a* values in which the FM and AFM phases have the same energy.

In **Figure 5a** total energy results for tetragonal LSMO (i.e. with squared substrate) for the four magnetic orderings are reported, as a function of the planar lattice parameter. FM and A-type AFM orderings tightly compete within the examined structural range; the others are much higher in energy and can be discarded. The A-AFM prevails in most of the examined *a* range, and is enhanced by increasing *a,* which corresponds to epitaxial tensile strain. On the other hand, FM is strengthened by compressive strain, and sets in for $a$ < 3.87 Å. In their respective equilibrium structures (corresponding both to $a_0 \approx$ 3.95-3.96 Å), FM and A-AFM orders differ by an energy of 25 meV/f.u. The interpretation of the FM vs. A-AFM competition is enlightened by the calculated $c/a$ ratio (**Figure 5b)** which decreases for increasing $a$. Importantly, for any given $a$ value, $c/a$ is always smaller (by a factor ≈ 0.01 on average) for the A-AFM phase than for the FM phase. The smaller $c/a$ ratio reflects a higher anisotropy factor (Figure 5b), defined as a mean square deviation of the cell parameters from their average. Notice that anisotropy vanishes at $a$ =3.815 Å and $a$ =3.78 Å for FM and A-AFM order, respectively, corresponding to the 3D cubic structures, while $a_0$ corresponds to a large (≈5%) anisotropy.

The results for orthorhombic LSMO (i.e. with rectangular substrate) are shown in **Figure 5c** and **5d**. The general trend observed in the calculations is that for $b/a$ < 1 the FM phase gains stability over the A-AFM. The turn-around occurs at $b/a$ =0.95 for $a$ =4 Å, and the $b/a$ value approaches 1 as $a$ is decreased. For $b/a$ > 1, on the other hand, the A-AFM phase is further strengthened with respect to the FM phase.

## IV. DISCUSSION

The known mechanism of magnetoelectric coupling are four: iron migration, charge accumulation/depletion, strain mediated and exchange mediated. Since the BTO substrate is not magnetic, the last case can be excluded. The fact that XAS lineshape does not change rules out ion migration as a possible cause: the chemical environment of Mn remains the same. Charge effects can also be excluded, since the detected XMCD signal is invariant for electric polarization reversal. Therefore, strain-mediated magnetoelectric coupling is the only possible explanation of the observed phenomena. In the following, the experimental results are interpreted according to this view, supported by the simulations described in the previous section.

LSMO in the 20-35 % doping range is FM in the bulk whereas for epitaxially grown strained thin films the magnetic ordering may be different[39–42]. The magnetic ordering in LSMO is the result of the interplay between super-exchange and double-exchange interactions. The first is mediated by $t_{2g}$ orbitals and favors AFM ordering, while the latter by $e_g$ orbitals ($z^2$ or $x^2$-$y^2$) and favors FM ordering. In bulk, the dominant contribution of Mn $e_g$ coupling (via double-exchange with O(p) orbitals) in both planar and longitudinal directions favors spin-pairing in the three directions and overall, FM ordering. An applied strain along a given direction determines an anisotropic redistribution of the $e_g$ levels. In-plane tensile strain would cause a depletion of $z^2$ orbitals and charge accumulation in $x^2$-$y^2$ orbitals, with a consequent strengthening of FM ordering in-plane, and AFM super-exchange interactions prevailing across different planes, along the orthogonal direction.[40,43] The results of our simulations are consistent with this picture: the *c/a* ratio is the key parameter governing $e_g$ charge anisotropy, and consequently the magnetic ordering. Higher values of *a* correspond to smaller *c/a* values and higher anisotropy of the unit cell, pushing the system towards A-AFM ordering. FM ordering counteracts the effect of this charge redistribution resulting in equilibrium *c/a* values systematically larger than those for the A-AFM phase. The interpretation of the results for $b/a \neq 1$ are consistent with the results for *c/a* and anisotropy factor (see **Figure 5d** for the specific case a=3.89 Å): the decrease of *b/a* below unity increases the equilibrium *c/a* value, and in turn, decreases the anisotropy; this mechanism stabilizes the FM phase against the A-AFM. In tetragonal LSMO the turnaround occurred for *c/a* greater than 0.95-0.96. This behavior is substantially maintained even for the orthorhombic structures. Our analysis is also consistent with the results of previous computational studies of phase transitions induced in LSMO by compressive substrates[44].

The experimental results can now be interpreted: when BTO is in the T phase and unbiased, the in-plane tensile strain imposed on the LSMO film is large, favoring AFM order; in this situation FM ordering is hindered also by the disorder caused by the presence of multi-domains (both in- and out-of-plane) in the BTO substrate. When BTO is polarized out-of-plane, this disorder is reduced, and a cubic lattice is formed at the interface; this is accompanied with a reduction of the tensile strain imposed on the LSMO film, which favors the appearance of FM ordering. Evidently, this effect dominates over the loss of in-plane anisotropy, which acts contrariwise. The XMCD signal observed in this case is however very small (around 1%), indicating the competition between the effects of these subtle distortions. It is also important to note that the Curie temperature ($T_C$) of a tensile-strained LSMO film is reduced with respect to the bulk value[26], and hence the system could be close to the paramagnetic transition, with a reduced magnetization.

When BTO is in the O phase, we could expect the polarization vector to point 45° from the film plane, resulting in the formation of a (pseudo-) rectangular lattice at the interface. Even in this case there is a competition between the small increase of the substrate lattice area and the uniaxial deformation in determining the anisotropy of the LSMO unit cell. Our measurements indicate that the second effect is overcoming the first one, resulting in an overall stronger FM order of the LSMO film with respect to the T phase. This may be due even to the lowering of the temperature. Indeed, the intensity of the dichroic signal is reduced with the increase of temperature already in the O phase, vanishing in the case of polarized BTO at a temperature close to 320 ± 15 K (see Figure 4b), which can be assumed as the $T_C$ of the polarized case, a value smaller than that of bulk LSMO ($T_C$ = 369 K[25]).

Finally, when BTO transforms from the O to the R phase, the uniaxial deformation imposed to the LSMO film disappears, but the average tensile strain is not relieved. This favors the AFM ordering against FM, and indeed no XMCD was measured in this case.  It results therefore that the structural transition between R and O phases in BTO substrate leads to a magnetic transition from AFM to FM in LSMO thin film whose origin is strain-driven.

It is interesting to notice that although the changes in the BTO crystal parameters are lower than 1%, the corresponding magnetic effect on LSMO is sizeable. This once again confirms the strong interplay between the orbital and spin degrees of freedom in this transition metal oxide, and how the strain crucially affects the competition between FM and AFM orderings. XMCD cannot provide the experimental evidence of the existence of a AFM ordering; however, orbital anisotropy was already demonstrated for LSMO epitaxial film grown on substrates with a lower mismatch[39,45,46], so it is expected in this case too, also taking into account the insulating behavior observed from transport measurements in the R phase (see Figure 4).

Another aspect to be understood is the smallness of the XMCD signal observed compared to the value around 20% in the case of unstrained LSMO[36] (corresponding to a magnetization of 3.5 $\mu_B$/Mn[38]). As mentioned above, tensile stress in LSMO epitaxial films is known to decrease $T_C$, which implies that magnetization is severely reduced. Furthermore, our simulations show that AFM and FM orderings are in tight energetic competition for a wide range of lattice parameters. Several experimental and theoretical works (summarized in the review of Dagotto, Hotta and Moreo[47]) have been demonstrating the tendency of manganites to form an inhomogeneous state in which AFM and FM phases coexist, especially at the boundary of the phase diagram. Hence the changes of the Mn XMCD signal can be attributed to a variation of the FM fraction in the LSMO film, which is modulated by the substrate-induced strain. The smallness of this signal indicates that, in agreement with the simulations, the system would preferentially be AFM, but for some distortions of the substrate lattice it is pushed to the FM transition.

## V. CONCLUSIONS

We employed XMCD to study the magnetic response of a 30 u.c. LSMO film deposited on BTO, and its dependence on the crystal structure of the substrate. The results show that the magnetic ordering of LSMO is extremely sensitive to the small distortions induced by the structural phase transitions of the substrate. In the case of pristine BTO substrate, with a large majority of in-plane BTO domains, magnetic dichroism is observed for the (intermediate) BTO O phase, whereas no magnetic dichroism is detected for the T (high temperature) and R (low temperature) phases. After setting by means of an external bias an out-of-plane polarization of the substrate, i.e. aligning the majority of BTO domains to the out-of-plane direction, magnetic dichroism is measured in the LSMO film at room temperature (BTO in the tetragonal phase).

These observations show that fine engineering of the interfacial strain is a suitable way towards electric control of the magnetic state in manganites. The subtle interplay between overall strain and uniaxial in-plane deformation governs the competition between FM and AFM orderings as reflected also by the ab-initio calculations. The small changes in the LSMO epitaxial strain determined by changing the ratio between in-plane and out-of-plane domains in BTO substrate determine the transition between antiferromagnetism and ferromagnetism of the film.

## VI. EXPERIMENTAL AND THEORETICAL METHODS

A thin film of 30 u.c. (≈12 nm) of $La_{0.65}Sr_{0.35}MnO_3$ has been deposited by molecular beam epitaxy (MBE) on unpoled (100) BTO substrate from in an ozone atmosphere with $p = 5\times10^{-7}$ mbar, with the substrate kept at 1000 K. RHEED (reflection high-energy electron diffraction) assisted shuttered deposition developed by D. Schlom group[48] allowed to artificially repeat LSMO perovskite structure ($AO$-$BO_2$, $A$ being $La_{0.65}Sr_{0.35}$ and $B$ Mn respectively), with a control of the stoichiometry of the film during deposition.

XAS and XMCD measurements at Mn $L_{2,3}$ edges were performed at Advanced Photoelectric Effect beamline high energy branch (APE-HE)[29]. Total electron yield (TEY) detection system was used, allowing a probing depth through LSMO layer around 8 nm. Since the film is 12 nm thick, XMCD measurements probe a significant fraction of the volume of the film. A "magnetically dead layer" is known to form at the substrate/LSMO interface, especially in the presence of a high strain. This interfacial region is beyond the probing depth of the measurements here presented. Absorption measurements have been taken in circular polarization, with the sample at 45° with respect to the incident beam. To minimize possible artifacts, alternating magnetic field pulses of +200 and -200 Oe have been applied in the plane of the sample surface at each measured point of the absorption spectra; the difference between the two resulting curves gives the dichroic signal of LSMO layer. The Sample was cooled down to 100 K through liquid nitrogen cooling system, and heated up to room temperature by a local heater. A thermocouple placed behind the sample holder allowed controlling the local temperature of the sample.

The sample was first characterized by XAS and XMCD with the BTO substrate in the pristine state. Then, the sample was capped with a thin (≈2 nm) gold layer, removed from the analysis chamber and mounted on a specific sample holder that allows to set the out-of-plane polarization of the BTO substrate inside the analysis chamber (see Figure 3d). A MgO slab 0.5 mm thick was inserted under the sample to avoid electric contact between top and bottom of the sample. An electric bias up to 500 V could be applied with a Keithley 6485 Picoammeter/Voltage Source, leading to a net polarization of the substrate in the out-of-plane direction, as confirmed by current vs. voltage curves ($I$-$V$, see supporting information) and XRD characterizations. After setting the out-of-plane polarization, the sample was reintroduced in the analysis chamber and the XMCD characterization in temperature was repeated with the BTO polarized out-of-plane. For comparison between the "up" and "down" cases, the substrate was polarized *in-situ* right before the XAS/XMCD measurements and the effect of the polarization switching was immediately checked with the acquisition of an *I-V* curve (see also the Supplementary Material file[33]).

A second sample was grown in the exact same condition, but without any gold capping layer, and its structural and transport properties were studied. XRD measurements in Bragg-Brentano geometry were performed with a PANalytical's EMPYREAN instrument[30] with Cu-$K_\alpha$ radiation at room temperature, i.e. with BTO in tetragonal phase. In the High-Intensity mode the incident radiation is not monochromatic (Figure 1a). High-Resolution XRD measurements were obtained in a double-axis configuration, using a 4-bounce Ge(220) monochromator to select only the Cu-$K_{\alpha 1}$ line (Figure 4). The resistance of the LSMO film for different temperatures was measured in a four-probe Van der Pauw configuration, with gold electrical contacts placed on LSMO film surface.

First-Principles calculations were performed using density-functional theory within generalized-gradient spin-density approximation, as implemented in the QUANTUM ESPRESSO code[49]. For our calculations we employed a basis set of plane waves and ultrasoft pseudopotentials with cut-off energies of 40 Ryd, , 4x4x4 k-point grid (corresponding to 32 ab-initio k points in the Irreducible Brillouin Zone) and Gaussian smearing of 0.005 Ryd. Fully relaxed 20-atom supercells were used for all the examined magnetic orderings; doping was treated by actual atomic substitutions.

**Associated content:** Supporting information. Electrical characterization of the BTO ferroelectric substrate (figures S1 and S2)

**Corresponding author:** *Federico Motti, email motti@iom.cnr.it

**Acknowledgments**

This work has been performed in the framework of the Nanoscience Foundry and Fine Analysis (NFFA-MIUR Italy Progetti Internazionali) facility. A.F. acknowledges MIUR (Italian Ministry for University and Research) for funding under Project PON04a2_00490 M2M "NETERGIT") and computational support of the CRS4 Computing Center (Piscina Manna, Pula, Italy), and PRACE (Project "UNWRAP").

**References**

[1] H. Ohno, M.D. Stiles, and B. Dieny, Proc. IEEE **104**, 1782 (2016).

[2] A. Ney, C. Pampuch, R. Koch, and K.H. Ploog, Nature **425**, 485 (2003).

[3] F. Matsukura, Y. Tokura, and H. Ohno, Nat. Nanotechnol. **10**, 209 (2015).

[4] M. Bibes and A. Barthélémy, Nat. Mater. **7**, 425 (2008).

[5] L.W. Martin and R. Ramesh, Acta Mater. **60**, 2449 (2012).

[6] N.A. Hill, J. Phys. Chem. B **104**, 6694 (2000).

[7] G. Vinai, A. Khare, D.S. Rana, E. Di Gennaro, B. Gobaut, R. Moroni, A.Y. Petrov, U. Scotti Di Uccio, G. Rossi, F. Miletto Granozio, G. Panaccione, and P. Torelli, APL Mater. **3**, 116107 (2015).

[8] R.-C. Peng, J.-M. Hu, K. Momeni, J.-J. Wang, L.-Q. Chen, and C.-W. Nan, Sci. Rep. **6**, 27561 (2016).

[9] V. Garcia, M. Bibes, and A. Barthélémy, Comptes Rendus Phys. **16**, 168 (2015).

[10] H.J. Zhao, W. Ren, Y. Yang, J. Íñiguez, X.M. Chen, and L. Bellaiche, Nat. Commun. **5**, 4021

(2014).

[11] G. Radaelli, D. Petti, E. Plekhanov, I. Fina, P. Torelli, B.R. Salles, M. Cantoni, C. Rinaldi, D. Gutiérrez, G. Panaccione, M. Varela, S. Picozzi, J. Fontcuberta, and R. Bertacco, Nat. Commun. **5**, 3404 (2014).

[12] M. Sperl, P. Torelli, F. Eigenmann, M. Soda, S. Polesya, M. Utz, D. Bougeard, H. Ebert, G. Panaccione, and C.H. Back, Phys. Rev. B **85**, 184428 (2012).

[13] C.H. Ahn, J.-M. Triscone, and J. Mannhart, Nature **424**, 1015 (2003).

[14] H.Y. Hwang, Y. Iwasa, M. Kawasaki, B. Keimer, N. Nagaosa, and Y. Tokura, Nat. Mater. **11**, 103 (2012).

[15] E. Arenholz, G. van der Laan, F. Yang, N. Kemik, M.D. Biegalski, H.M. Christen, and Y. Takamura, Appl. Phys. Lett. **94**, 72503 (2009).

[16] W. Eerenstein, M. Wiora, J.L. Prieto, J.F. Scott, and N.D. Mathur, Nat. Mater. **6**, 348 (2007).

[17] J. Heidler, C. Piamonteze, R. V Chopdekar, M.A. Uribe-Laverde, A. Alberca, M. Buzzi, A. Uldry, B. Delley, C. Bernhard, and F. Nolting, Phys. Rev. B **91**, 24406 (2015).

[18] A. Alberca, C. Munuera, J. Azpeitia, B. Kirby, N.M. Nemes, A.M. Perez-Muñoz, J. Tornos, F.J. Mompean, C. Leon, J. Santamaria, and M. Garcia-Hernandez, Sci. Rep. **5**, 17926 (2016).

[19] M.K. Lee, T.K. Nath, C.B. Eom, M.C. Smoak, and F. Tsui, Appl. Phys. Lett. **77**, 3547 (2000).

[20] R. V Chopdekar, J. Heidler, C. Piamonteze, Y. Takamura, A. Scholl, S. Rusponi, H. Brune, L.J. Heyderman, and F. Nolting, Eur. Phys. J. B **86**, 241 (2013).

[21] C. Thiele, K. Dörr, O. Bilani, J. Rödel, and L. Schultz, Phys. Rev. B **75**, 54408 (2007).

[22] C.A.F. Vaz, F.J. Walker, C.H. Ahn, and S. Ismail-Beigi, J. Phys. Condens. Matter **27**, 123001 (2015).

[23] G.H. Kwei, A.C. Lawson, S.J.L. Billinge, and S.W. Cheong, J. Phys. Chem. **97**, 2368 (1993).

[24] T.H.E. Lahtinen and S. Van Dijken, Appl. Phys. Lett. **102**, 112406 (2013).

[25] J. Hemberger, A. Krimmel, T. Kurz, H. Krug von Nidda, V.Y. Ivanov, A.A. Mukhin, A.M. Balbashov, and A. Loidl, Phys. Rev. B **66**, 94410 (2002).

[26] A.-M. Haghiri-Gosnet and J.-P. Renard, J. Phys. D. Appl. Phys. **36**, R127 (2003).

[27] F. Tsui, M.C. Smoak, T.K. Nath, and C.B. Eom, Appl. Phys. Lett. **76**, 2421 (2000).

[28] Z. Fang and K. Terakura, J. Phys. Soc. Japan **70**, 3356 (2001).

[29] G. Panaccione, I. Vobornik, J. Fujii, D. Krizmancic, E. Annese, L. Giovanelli, F. MacCherozzi, F. Salvador, A. De Luisa, D. Benedetti, A. Gruden, P. Bertoch, F. Polack, D. Cocco, G. Sostero, B. Diviacco, M. Hochstrasser, U. Maier, D. Pescia, C.H. Back, T. Greber, J. Osterwalder, M. Galaktionov, M. Sancrotti, and G. Rossi, Rev. Sci. Instrum. **80**, 43105 (2009).

[30](n.d.).

[31] J.-L. Maurice, F. Pailloux, A. Barthélémy, O. Durand, D. Imhoff, R. Lyonnet, A. Rocher, and J.-P. Contour, Philos. Mag. **83**, 3201 (2003).

[32] C. Adamo, X. Ke, H.Q. Wang, H.L. Xin, T. Heeg, M.E. Hawley, W. Zander, J. Schubert, P. Schiffer, D.A. Muller, L. Maritato, and D.G. Schlom, Appl. Phys. Lett. **95**, 112504 (2009).

[33]See supplementary material at [url] for a reciprocal space map around the 103 Bragg reflection, further RHEED data and details on BTO ferroelectric switching.

[34] M. Abbate, F.M.F. de Groot, J.C. Fuggle, A. Fujimori, O. Strebel, M.F. Lopez, M. Domke, G. Kaindl, G.A. Sawatzky, M. Takano, Y. Takeda, H. Eisaki, and S. Uchida, Phys. Rev. B **46**, 4511 (1992).

[35] A. Tebano, C. Aruta, P.G. Medaglia, F. Tozzi, G. Balestrino, A.A. Sidorenko, G. Allodi, R. De Renzi, G. Ghiringhelli, C. Dallera, L. Braicovich, and N.B. Brookes, Phys. Rev. B - Condens. Matter Mater. Phys. **74**, 1 (2006).

[36] G. Shibata, K. Yoshimatsu, E. Sakai, V.R. Singh, V.K. Verma, K. Ishigami, T. Harano, T. Kadono, Y. Takeda, T. Okane, Y. Saitoh, H. Yamagami, A. Sawa, H. Kumigashira, M. Oshima, T. Koide, and A. Fujimori, Phys. Rev. B **89**, 235123 (2014).

[37] T. Koide, H. Miyauchi, J. Okamoto, T. Shidara, T. Sekine, T. Saitoh, A. Fujimori, H. Fukutani, M. Takano, and Y. Takeda, Phys. Rev. Lett. **87**, 246404 (2001).

[38] A. Urushibara, Y. Moritomo, T. Arima, A. Asamitsu, G. Kido, and Y. Tokura, Phys. Rev. B **51**, (1995).

[39] D. Pesquera, A. Barla, M. Wojcik, E. Jedryka, F. Bondino, E. Magnano, S. Nappini, D. Gutiérrez, G. Radaelli, G. Herranz, F. Sánchez, and J. Fontcuberta, Phys. Rev. Appl. **6**, 34004 (2016).

[40] G. Colizzi, A. Filippetti, F. Cossu, and V. Fiorentini, Phys. Rev. B **78**, 235122 (2008).

[41] F. Cossu, U. Schwingenschlögl, G. Colizzi, A. Filippetti, and V. Fiorentini, Phys. Rev. B **87**, 214420 (2013).

[42] A.M. Haghiri-Gosnet, J. Wolfman, B. Mercey, C. Simon, P. Lecoeur, M. Korzenski, M. Hervieu, R. Desfeux, and G. Baldinozzi, J. Appl. Phys. **88**, 4257 (2000).

[43] G. Colizzi, A. Filippetti, and V. Fiorentini, Phys. Rev. B **76**, 64428 (2007).

[44] G. Colizzi, A. Filippetti, F. Cossu, and V. Fiorentini, Eur. Phys. J. B **70**, 343 (2009).

[45] H. Boschker, J. Kautz, E.P. Houwman, W. Siemons, D.H.A. Blank, M. Huijben, G. Koster, A. Vailionis, and G. Rijnders, Phys. Rev. Lett. **109**, 157207 (2012).

[46] C. Aruta, G. Ghiringhelli, V. Bisogni, L. Braicovich, N.B. Brookes, A. Tebano, and G. Balestrino, Phys. Rev. B **80**, 14431 (2009).

[47] E. Dagotto, T. Hotta, and A. Moreo, Phys. Rep. **344**, 1 (2001).

[48] J.H. Haeni, C.D. Theis, and D.G. Schlom, J. Electroceramics **4**, 385 (2000).

[49] P. Giannozzi, S. Baroni, N. Bonini, M. Calandra, R. Car, C. Cavazzoni, D. Ceresoli, G.L. Chiarotti, M. Cococcioni, I. Dabo, A.D. Corso, S. Fabris, G. Fratesi, S. de Gironcoli, R. Gebauer, U. Gerstmann, C. Gougoussis, A. Kokalj, M. Lazzeri, L. Martin-Samos, N. Marzari, F. Mauri, R. Mazzarello, S. Paolini, A. Pasquarello, L. Paulatto, C. Sbraccia, S. Scandolo, G. Sclauzero, A.P. Seitsonen, A. Smogunov, P. Umari, and R.M. Wentzcovitch, J. Phys. Condens. Matter **21**, 395502 (2009).